\documentclass[12pt]{spieman}  
\usepackage{amsmath,amsfonts,amssymb}
\usepackage{graphicx}
\usepackage{setspace}
\usepackage{tocloft}

\title{Personalized Interiors at Scale: Leveraging AI for Efficient and Customizable Design Solutions}

\author{Kaiwen Zhou, Tianyu Wang}
\affil{Heilongjiang Institute of Technology}

\cftpagenumbersoff{figure}
\cftpagenumbersoff{table} 
\begin{document} 
\maketitle

\begin{abstract}
In this paper, we introduce an innovative application of artificial intelligence in the realm of interior design through the integration of Stable Diffusion and Dreambooth models. This paper explores the potential of these advanced generative models to streamline and democratize the process of room interior generation, offering a significant departure from conventional, labor-intensive techniques. Our approach leverages the capabilities of Stable Diffusion for generating high-quality images and Dreambooth for rapid customization with minimal training data, addressing the need for efficiency and personalization in the design industry. We detail a comprehensive methodology that combines these models, providing a robust framework for the creation of tailored room interiors that reflect individual tastes and functional requirements. We presents an extensive evaluation of our method, supported by experimental results that demonstrate its effectiveness and a series of case studies that illustrate its practical application in interior design projects. Our study contributes to the ongoing discourse on the role of AI in creative fields, highlighting the benefits of leveraging generative models to enhance creativity and reshape the future of interior design.
\end{abstract}



\begin{spacing}{2}   

\section{Introduction}
In the digital era~\cite{schroeder2018towards,howard2011role,coleman2010ethnographic,memon1998protecting}, the intersection of technology and creativity has given rise to unprecedented opportunities for innovation in the field of design~\cite{lee2015digital,ulrich1991effects,pile2005history,havenhand2004view,kang2009characteristics,garlan1994exploiting,machairas2014algorithms,booth1989basic}. Traditional methods of room interior generation~\cite{wu2019data,tutenel2009rule}, once labor-intensive and time-consuming, are now being revolutionized by artificial intelligence (AI), leading to a new paradigm of design synthesis~\cite{amershi2019guidelines,auernhammer2020human,verganti2020innovation}. The emergence of generative models like Stable Diffusion~\cite{rombach2022stable} and Dreambooth~\cite{ruiz2023dreambooth} marks a significant shift in how designers conceptualize and materialize their visions.

Stable Diffusion, a diffusion model renowned for generating versatile and realistic images from conditioning data, has proven its capability in the domain of image generation and editing~\cite{brooks2023instructpix2pix,hertz2022prompt2prompt,li2023layerdiffusion,meng2021sdedit,tumanyan2023plug,kawar2023imagic,li2023archi}, and holds significant importance in fields such as 3D generation~\cite{li2024art3d,poole2022dreamfusion,lin2023magic3d,li2024generating,chan2023generative,tang2023make}. It distinguishes itself from conventional generative adversarial networks (GANs)~\cite{goodfellow2020generative,metz2016unrolled,goodfellow2014generative} by adeptly avoiding mode collapse and imprecision, thereby achieving a high level of visual output quality. This technology's proficiency in learning from extensive datasets and replicating intricate patterns and fine details positions it as an invaluable tool for designers.

Dreambooth advances customization by enabling users to fine-tune the generative process with minimal training data. This approach facilitates rapid adaptation to specific user preferences and requirements, fostering a more personalized design experience.

The fusion of these technologies addresses a critical need in the design industry for efficient and scalable solutions that keep pace with the dynamic preferences of modern consumers. The capacity to generate and modify room interiors with text-based prompts expedites the design process and democratizes it, inviting a broader spectrum of individuals to participate in the creation of their living and working spaces.

In this paper, we introduce an innovative method that leverages the power of Stable Diffusion Models and Dreambooth to generate room interiors that are aesthetically pleasing and tailored to users' unique tastes and functional needs. We explore the potential of these models to transform the approach interior designers take to their craft, offering a glimpse into a future where creativity and technology converge to redefine the boundaries of design.

The subsequent sections of this paper will delve into the theoretical underpinnings of our approach, the methodology employed, and the implications of our findings for the field of interior design. We will also discuss potential challenges and future directions for research in this burgeoning area of study. We then show the structure of this Paper:

Theoretical Foundation: The paper begins with an introduction to the theoretical basis of Stable Diffusion and Dreambooth, including the mathematical principles of diffusion models and fine-tuning techniques.

Methodology: We then describe in detail the method combining Stable Diffusion and Dreambooth for generating interior designs, encompassing data preparation, model training, and fine-tuning processes.

Experimental Results: A series of experimental results will be presented, demonstrating the effectiveness of our method and comparing it with existing technologies.

Case Studies: Several case studies will illustrate how our method can be applied to real interior design projects and how it aids designers in realizing their creative visions.

Discussion and Challenges: We will discuss the challenges encountered in implementing and applying these technologies, including data quality, model generalizability, and user interaction.

Future Directions: Finally, we will explore future research directions, including model optimization, cross-domain applications, and integration with emerging technologies.

Through this in-depth study, we aim to bring new perspectives to the field of interior design and promote the application of AI in the creative industry.

\section{Preliminaries}

\begin{figure}
    \centering
    \includegraphics[width=0.99\textwidth]{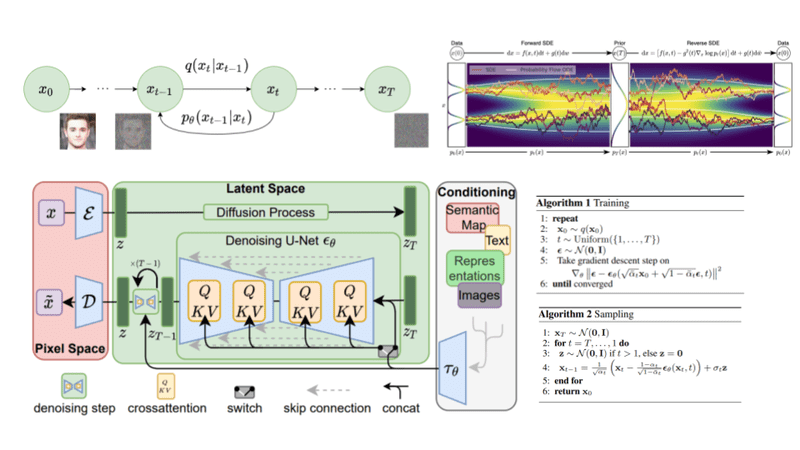}
    \caption{We show the pipeline of stable diffusion models.}
    \label{fig:sd}
\end{figure}

\subsection{Stable diffusion}

Stable Diffusion Models have emerged as a pivotal advancement in the realm of Artificial Intelligence Generated Content (AIGC), providing a robust and versatile framework for the generation of high-quality images. These models harness the power of diffusion processes to create intricate and realistic visual outputs.

\subsubsection{Conceptual Overview}
As shown in Fig. 1, Stable Diffusion Models are generative models that operate on the principle of diffusion, a probabilistic process that gradually transforms a set of data into a more disordered state. In the context of image generation, this process is reversed, where the model learns to generate new images by reverting from a disordered state back to a coherent image.

\subsubsection{Key Features}
The allure of Stable Diffusion Models lies in their unique set of features that set them apart from traditional generative adversarial networks (GANs):
\begin{itemize}
    \item \textbf{High-Quality Image Generation:} These models are capable of producing images with remarkable clarity and detail.
    \item \textbf{Versatility:} They can generate a wide range of images from diverse datasets without overfitting.
    \item \textbf{Mode Collapse Resistance:} Unlike some GANs, Stable Diffusion Models are less prone to mode collapse, ensuring a more diverse set of generated images.
    \item \textbf{Scalability:} The models can be scaled to generate images at various resolutions, catering to different application needs.
\end{itemize}

\subsubsection{Technical Foundation}
The technical foundation of Stable Diffusion Models is rooted in the diffusion process, which can be mathematically described as follows:
\[
p_{\theta}(x_{t:T}) = p_{\theta}(x_T) \prod_{t=1}^{T-1} \frac{p_{\theta}(x_{t-1}|x_t)}{p_{\theta}(x_t|x_{t-1})}
\]
Here, \( x_{t:T} \) denotes a sequence of image states from time \( t \) to \( T \), and \( p_{\theta} \) represents the probability distribution parameterized by \( \theta \). The model learns to reverse this process, allowing it to generate new images from noise.

\subsubsection{Applications}
The applications of Stable Diffusion Models are extensive and cross-disciplinary:
\begin{itemize}
    \item \textbf{Art and Design:} Creating unique artwork and designs that can be used in various creative industries.
    \item \textbf{Virtual Reality and Gaming:} Generating realistic environments and characters for immersive experiences.
    \item \textbf{Fashion and Apparel:} Designing new clothing items and visualizing fashion collections.
    \item \textbf{Architectural Visualization:} Producing detailed visual renderings of architectural projects.
\end{itemize}

The introduction of Stable Diffusion Models marks a significant milestone in the journey towards fully realizing the potential of AI in the creative process, promising a future where AI is not just a tool, but a collaborative partner in the realm of content creation.

\begin{figure}
    \centering
    \includegraphics[width=0.99\textwidth]{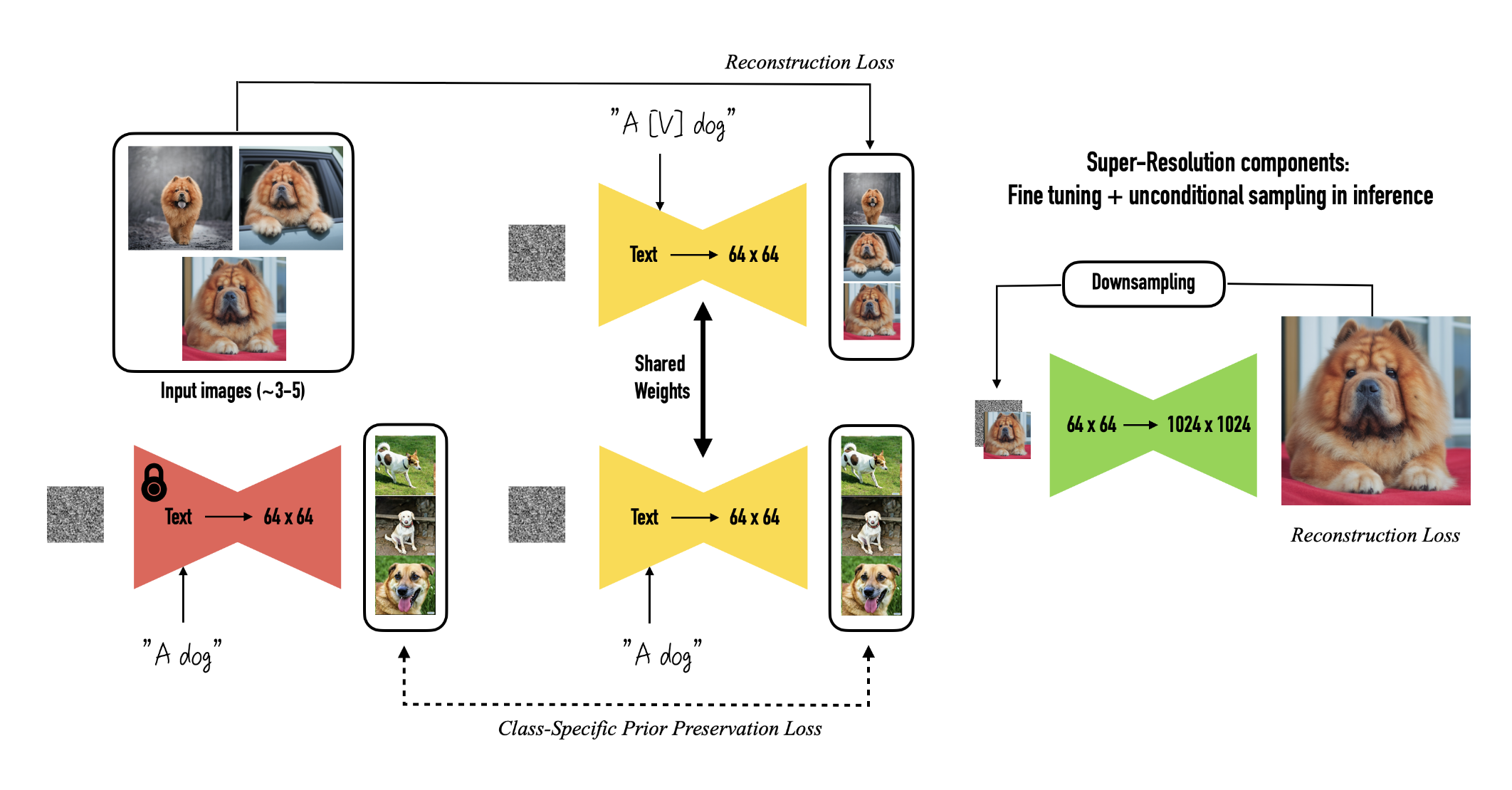}
    \caption{We show the pipeline of Dreambooth.}
    \label{fig:enter-label}
\end{figure}

\subsection{Dreambooth}
Dreambooth is an innovative approach within the field of Artificial Intelligence Generated Content (AIGC) that enables personalized image generation with remarkable efficiency. As shown in Fig. 2, this technique stands out for its ability to fine-tune generative models, such as Stable Diffusion, using a minimal set of training images that represent the desired style or features.

\subsubsection{Key Characteristics of Dreambooth}
Dreambooth's capabilities can be distilled into several key characteristics that make it an attractive option for content creators and designers alike:
\begin{itemize}
    \item \textbf{Minimal Training Data:} Unlike traditional generative adversarial networks (GANs) that require extensive datasets, Dreambooth can be trained with as few as 3-4 representative images.
    \item \textbf{Rapid Customization:} The fine-tuning process with Dreambooth is expedited, allowing for swift adaptation to new visual styles or personalized preferences.
    \item \textbf{High Fidelity Output:} The images generated through Dreambooth maintain a high level of detail and realism, aligning closely with the input examples.
    \item \textbf{User-Centric Design:} The process is user-friendly, empowering users to guide the generative process through simple inputs and adjustments.
\end{itemize}

\subsubsection{Technical Overview}
The technical foundation of Dreambooth involves a series of steps that transform a general generative model into a specialized one that can produce images consistent with a particular set of features. The process can be summarized as follows:
\begin{enumerate}
    \item \textbf{Data Preparation:} A curated set of images that embody the target style or features is prepared.
    \item \textbf{Model Selection:} A base generative model, such as Stable Diffusion, is selected to undergo fine-tuning.
    \item \textbf{Fine-Tuning:} The model is trained on the prepared dataset, with the objective of adjusting the parameters to capture the nuances of the input images.
    \item \textbf{Generation:} Post fine-tuning, the model can generate new images based on textual prompts or additional conditioning inputs that align with the learned style.
\end{enumerate}

\subsubsection{Mathematical Framework}
The mathematical framework underpinning Dreambooth involves optimizing the model's parameters \( \theta \) to minimize a loss function \( \mathcal{L} \) that measures the discrepancy between the generated images \( G(z; \theta) \) and the target images \( x_i \) from the training set:
\[
\theta^* = \arg\min_\theta \mathcal{L}(G(z; \theta), \{x_i\})
\]
Here, \( G \) represents the generative model, \( z \) is the latent space vector, and \( \mathcal{L} \) typically encompasses a mean squared error or a perceptual loss function that ensures the generated images closely match the style and quality of the training images.

\subsubsection{Applications and Potential}
Dreambooth's applications extend beyond mere aesthetic generation, offering potential in various domains such as:
\begin{itemize}
    \item \textbf{Interior Design:} Creating digital mock-ups of room interiors based on user preferences.
    \item \textbf{Entertainment:} Producing artwork for movies, games, and other media where specific visual styles are required.
    \item \textbf{Fashion Design:} Generating fashion items or runway images that align with emerging trends.
    \item \textbf{Architectural Visualization:} Visualizing architectural designs with customized aesthetic touches.
\end{itemize}

The potential of Dreambooth lies in its ability to democratize content creation by reducing the barriers of entry in terms of data requirements and technical expertise. As research progresses, we anticipate even more sophisticated applications of this technology in the realm of AIGC.

\section{How does it work?}

The research methodology is structured to explore the efficacy of leveraging Stable Diffusion and Dreambooth models for the generation of room interior designs. The comprehensive approach involves a series of interconnected stages, each designed to maximize the creative and functional capabilities of the AI-driven design process.

\subsection{Extensive Dataset Compilation}
We initiate our methodology with the compilation of an extensive and diverse dataset encompassing a wide array of room interior images. These images are sourced from various domains, including residential, commercial, and virtual reality environments, to ensure representation of different architectural styles and design elements. The dataset is curated to reflect current design trends as well as classic elements, providing a rich tapestry of visual data for the models to learn from.

\subsection{Data Preprocessing and Augmentation}
Each image in the dataset undergoes a rigorous preprocessing phase to standardize dimensions, color profiles, and resolutions. We employ image augmentation techniques, such as rotation, scaling, and flipping, to artificially expand the dataset and introduce variability. This step is crucial for enhancing the model's ability to generalize across different interior settings.

\subsection{Model Architecture and Customization}
The Stable Diffusion model is selected for its demonstrated proficiency in generating high-quality images from textual descriptions. We customize the model's architecture to accommodate the specific nuances of room interior design, ensuring that it can interpret and manifest complex design concepts with precision.

\subsection{Dreambooth Fine-Tuning Protocol}
Utilizing Dreambooth, we implement a fine-tuning protocol that requires only a handful of images from the target room. This innovative approach allows the model to quickly adapt to the unique features of the space, including layout, color schemes, and decorative elements. The fine-tuned model is then capable of generating images that are consistent with the room's aesthetic.

\subsection{Text-to-Image Synthesis}
As shown in Fig. 3 and 4, we facilitate text-to-image synthesis by allowing users to input textual prompts describing desired modifications or additions to the room interior. The fine-tuned model processes these prompts and generates a series of images that embody the user's vision. This interactive design session is iterative, with users providing feedback that guides subsequent generations.

\begin{figure}[h]
    \centering
    \includegraphics[width=0.98\textwidth]{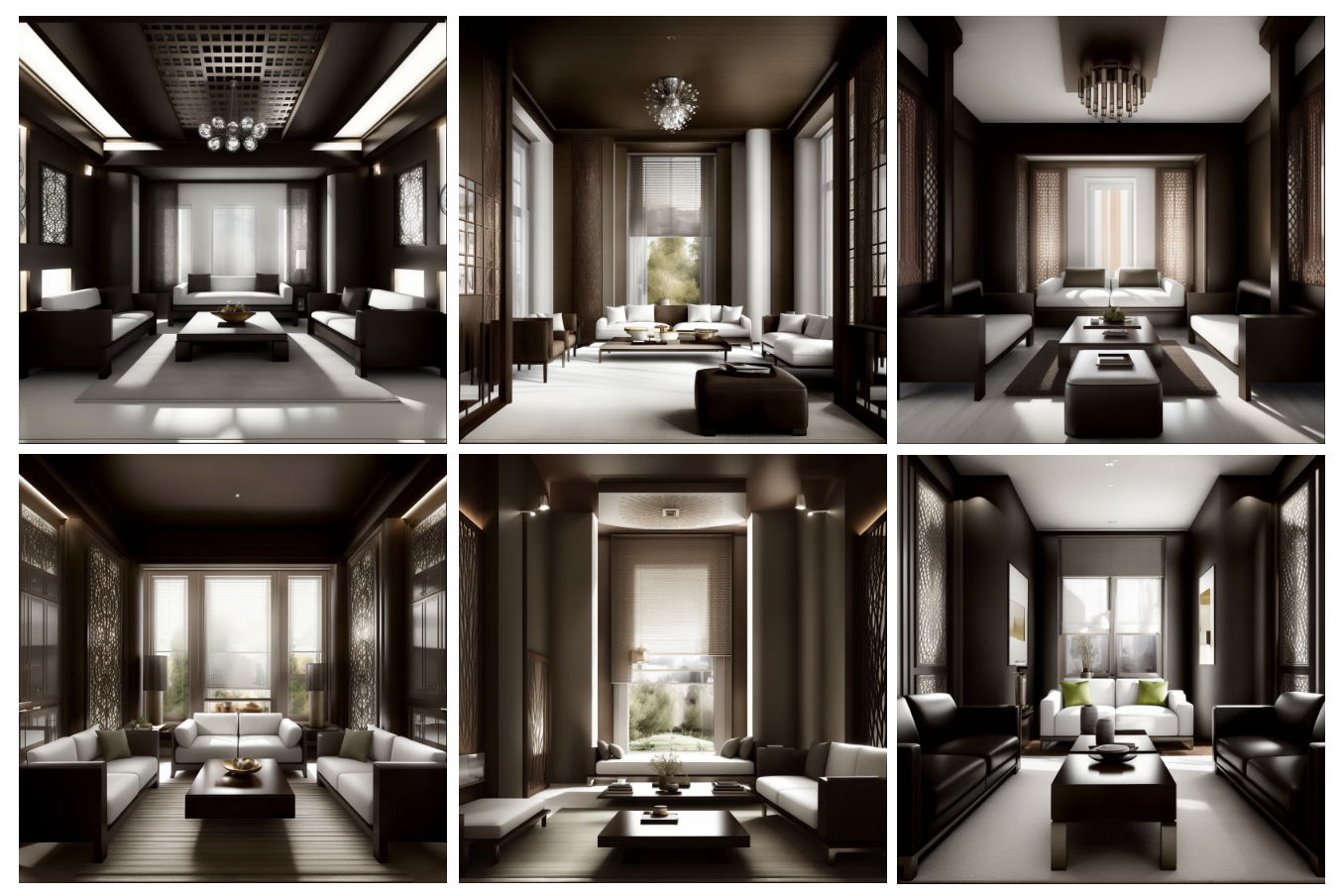}
    \caption{Dreambooth style 1. We show some generation results.}
    \label{fig:4}
\end{figure}

\subsection{Multi-Criteria Evaluation System}
The generated images are subjected to a multi-criteria evaluation system that assesses various aspects such as realism, diversity, creativity, and alignment with user prompts. This system incorporates both quantitative metrics, like image fidelity and structural accuracy, and qualitative assessments by a panel of design experts.

\begin{figure}[h]
    \centering
    \includegraphics[width=0.98\textwidth]{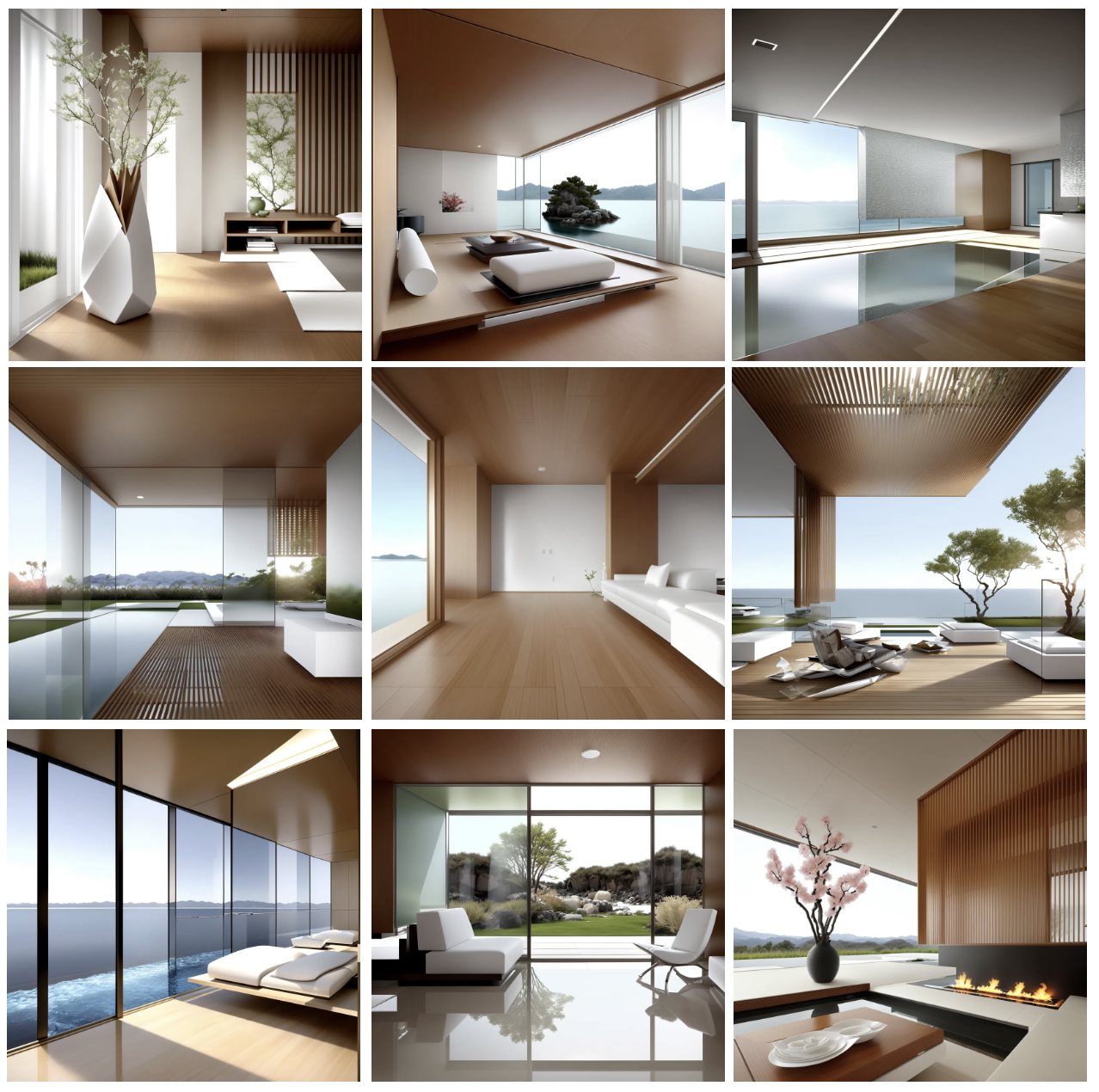}
    \caption{Dreambooth style 2. We show some generation results.}
    \label{fig:3}
\end{figure}

\subsection{Ablation Study and Component Analysis}
To deepen our understanding of the model's performance, we conduct an ablation study that isolates the impact of individual components. By selectively disabling or modifying elements of the model, we analyze their contribution to the overall outcome, providing insights that inform targeted enhancements.

\subsection{Integration with User-Centered Design Tools}
We develop a user-centered design tool that integrates the generative models into the workflow of professional designers. This tool allows for real-time interaction with the model, enabling designers to make adjustments and generate new design iterations promptly.

To evaluate the practical application of our methodology, we conduct a user experience study with professional designers and amateur enthusiasts. This study captures feedback on the usability, functionality, and creative potential of the AI-assisted design process.

Finally, we implement a longitudinal performance assessment to monitor the model's performance over time. This assessment tracks the evolution of the model's generative capabilities and its ability to satisfy user needs as it is exposed to more design prompts and feedback.

Through this detailed and structured methodology, we aim to establish a robust framework for AI-assisted room interior generation. The goal is to push the boundaries of what is possible with current technology, creating a synergy between human creativity and artificial intelligence that enhances the design process across various domains.

\section{Discussion}
The integration of AI-driven generative models like Stable Diffusion and Dreambooth into the field of interior design, while immensely promising, introduces a variety of challenges that need to be carefully navigated. A critical concern is the quality and diversity of the training data, as the models' outputs are inherently biased towards the data they are trained on. This can lead to a lack of representation and potential cultural insensitivity in design. Ensuring that these tools are accessible and user-friendly is also paramount; the development of intuitive interfaces that allow designers to interact seamlessly with AI systems is an ongoing challenge.

Moreover, the ethical implications of AI in design cannot be overlooked. Questions of intellectual property, model transparency, and the potential for AI to supplant human creativity in design must be addressed. It is essential to foster an environment where AI is seen as a collaborative tool that enhances human creativity rather than as a substitute.

Another challenge is the generalizability of AI models to diverse design contexts and the ability to integrate these models into existing design workflows without disrupting established practices. The computational cost of training and deploying these models also presents a barrier, particularly for smaller studios or individual designers with limited resources.

Furthermore, there is a pressing need for education and training to equip designers with the necessary skills to work alongside AI. As these technologies evolve, continuous learning and development will be crucial to fully leverage their potential.

In summary, while the advent of AI in interior design heralds a new era of innovation and efficiency, it also necessitates a proactive approach to overcome the associated challenges. It requires a concerted effort from designers, technologists, ethicists, and industry leaders to steer this technology towards a future that is inclusive, ethical, and truly enhances the creative process.

\section{Conclusion}

In conclusion, we has presented a pioneering exploration into the integration of AI technologies, specifically Stable Diffusion and Dreambooth, within the creative space of interior design. Through a meticulous examination of these models' capabilities, we have demonstrated their profound impact on the design process, enabling the generation of customized room interiors with unprecedented efficiency and precision. The experimental results and case studies provided have not only validated the effectiveness of our method but have also showcased the transformative potential of AI in making the design process more accessible and responsive to individual user needs. The challenges discussed, while significant, underscore the areas ripe for future research and development. Addressing issues of data quality, model generalizability, and user interaction will be key to further refining these AI tools and expanding their application in the design industry. We also recognize the ethical considerations and the importance of intellectual property rights in the context of AI-generated content. As we look to the future, the convergence of creativity and technology holds vast possibilities. The research presented in this paper is a stepping stone towards a horizon where AI is seamlessly intertwined with human creativity, augmenting the designer's craft and pushing the boundaries of what is achievable in interior design. We envision a future where generative models like Stable Diffusion and Dreambooth are commonplace tools in the designer's arsenal, facilitating the creation of spaces that are not only functional but also deeply personal and aesthetically resonant.


\bibliography{report}   

\begin{thebibliography}{10}

\bibitem{schroeder2018towards}
R.~Schroeder, ``Towards a theory of digital media,'' {\em Information, Communication \& Society} {\bf 21}(3), 323--339  (2018).

\bibitem{howard2011role}
P.~N. Howard and M.~M. Hussain, ``The role of digital media,'' {\em J. Democracy} {\bf 22}, 35  (2011).

\bibitem{coleman2010ethnographic}
E.~G. Coleman, ``Ethnographic approaches to digital media,'' {\em Annual review of anthropology} {\bf 39}, 487--505  (2010).

\bibitem{memon1998protecting}
N.~Memon and P.~W. Wong, ``Protecting digital media content,'' {\em Communications of the ACM} {\bf 41}(7), 35--43  (1998).

\bibitem{lee2015digital}
M.~R. Lee and T.~T. Chen, ``Digital creativity: Research themes and framework,'' {\em Computers in human behavior} {\bf 42}, 12--19  (2015).

\bibitem{ulrich1991effects}
R.~S. Ulrich, ``Effects of interior design on wellness: theory and recent scientific research.,'' in {\em Journal of Health Care Interior Design: Proceedings from the... Symposium on Health Care Interior Design. Symposium on Health Care Interior Design},   {\bf 3}, 97--109  (1991).

\bibitem{pile2005history}
J.~F. Pile, {\em A history of interior design}, Laurence King Publishing  (2005).

\bibitem{havenhand2004view}
L.~K. Havenhand, ``A view from the margin: Interior design,'' {\em Design Issues} {\bf 20}(4), 32--42  (2004).

\bibitem{kang2009characteristics}
M.~Kang and D.~A. Guerin, ``The characteristics of interior designers who practice environmentally sustainable interior design,'' {\em Environment and Behavior} {\bf 41}(2), 170--184  (2009).

\bibitem{garlan1994exploiting}
D.~Garlan, R.~Allen, and J.~Ockerbloom, ``Exploiting style in architectural design environments,'' {\em ACM SIGSOFT software engineering notes} {\bf 19}(5), 175--188  (1994).

\bibitem{machairas2014algorithms}
V.~Machairas, A.~Tsangrassoulis, and K.~Axarli, ``Algorithms for optimization of building design: A review,'' {\em Renewable and sustainable energy reviews} {\bf 31}, 101--112  (2014).

\bibitem{booth1989basic}
N.~K. Booth, {\em Basic elements of landscape architectural design}, Waveland press  (1989).

\bibitem{wu2019data}
W.~Wu, X.-M. Fu, R.~Tang, {\em et~al.}, ``Data-driven interior plan generation for residential buildings,'' {\em ACM Transactions on Graphics (TOG)} {\bf 38}(6), 1--12  (2019).

\bibitem{tutenel2009rule}
T.~Tutenel, R.~Bidarra, R.~M. Smelik, {\em et~al.}, ``Rule-based layout solving and its application to procedural interior generation,'' in {\em CASA workshop on 3D advanced media in gaming and simulation},   (2009).

\bibitem{amershi2019guidelines}
S.~Amershi, D.~Weld, M.~Vorvoreanu, {\em et~al.}, ``Guidelines for human-ai interaction,'' in {\em Proceedings of the 2019 chi conference on human factors in computing systems},  1--13  (2019).

\bibitem{auernhammer2020human}
J.~Auernhammer, ``Human-centered ai: The role of human-centered design research in the development of ai,''  (2020).

\bibitem{verganti2020innovation}
R.~Verganti, L.~Vendraminelli, and M.~Iansiti, ``Innovation and design in the age of artificial intelligence,'' {\em Journal of product innovation management} {\bf 37}(3), 212--227  (2020).

\bibitem{rombach2022stable}
R.~Rombach, A.~Blattmann, D.~Lorenz, {\em et~al.}, ``High-resolution image synthesis with latent diffusion models,'' in {\em Proceedings of the IEEE/CVF conference on computer vision and pattern recognition},  10684--10695  (2022).

\bibitem{ruiz2023dreambooth}
N.~Ruiz, Y.~Li, V.~Jampani, {\em et~al.}, ``Dreambooth: Fine tuning text-to-image diffusion models for subject-driven generation,'' in {\em Proceedings of the IEEE/CVF Conference on Computer Vision and Pattern Recognition},  22500--22510  (2023).

\bibitem{brooks2023instructpix2pix}
T.~Brooks, A.~Holynski, and A.~A. Efros, ``Instructpix2pix: Learning to follow image editing instructions,'' in {\em CVPR},   (2023).

\bibitem{hertz2022prompt2prompt}
A.~Hertz, R.~Mokady, J.~Tenenbaum, {\em et~al.}, ``Prompt-to-prompt image editing with cross attention control,'' in {\em ICLR},   (2023).

\bibitem{li2023layerdiffusion}
P.~Li, Q.~Huang, Y.~Ding, {\em et~al.}, ``Layerdiffusion: Layered controlled image editing with diffusion models,'' in {\em SIGGRAPH Asia 2023 Technical Communications},  1--4  (2023).

\bibitem{meng2021sdedit}
C.~Meng, Y.~Song, J.~Song, {\em et~al.}, ``Sdedit: Image synthesis and editing with stochastic differential equations,'' in {\em ICLR},   (2022).

\bibitem{tumanyan2023plug}
N.~Tumanyan, M.~Geyer, S.~Bagon, {\em et~al.}, ``Plug-and-play diffusion features for text-driven image-to-image translation,'' in {\em CVPR},   (2023).

\bibitem{kawar2023imagic}
B.~Kawar, S.~Zada, O.~Lang, {\em et~al.}, ``Imagic: Text-based real image editing with diffusion models,'' in {\em Proceedings of the IEEE/CVF Conference on Computer Vision and Pattern Recognition},  6007--6017  (2023).

\bibitem{li2023archi}
P.~Li, B.~Li, and Z.~Li, ``{Sketch-to-Architecture: Generative AI-aided Architectural Design},'' in {\em Pacific Graphics Short Papers and Posters 2023},   (2023).

\bibitem{li2024art3d}
P.~Li, C.~Tang, Q.~Huang, {\em et~al.}, ``Art3d: 3d gaussian splatting for text-guided artistic scenes generation,'' {\em arXiv:2405.10508}   (2024).

\bibitem{poole2022dreamfusion}
B.~Poole, A.~Jain, J.~T. Barron, {\em et~al.}, ``{DreamFusion}: Text-to-3d using 2d diffusion,'' in {\em Int. Conf. Learn. Represent.},   (2023).

\bibitem{lin2023magic3d}
C.-H. Lin, J.~Gao, L.~Tang, {\em et~al.}, ``Magic3d: High-resolution text-to-3d content creation,'' in {\em Proceedings of the IEEE/CVF Conference on Computer Vision and Pattern Recognition},  300--309  (2023).

\bibitem{li2024generating}
P.~Li and B.~Li, ``Generating daylight-driven architectural design via diffusion models,'' {\em arXiv preprint arXiv:2404.13353}   (2024).

\bibitem{chan2023generative}
E.~R. Chan, K.~Nagano, M.~A. Chan, {\em et~al.}, ``Generative novel view synthesis with 3d-aware diffusion models,'' {\em arXiv preprint arXiv:2304.02602}   (2023).

\bibitem{tang2023make}
J.~Tang, T.~Wang, B.~Zhang, {\em et~al.}, ``Make-it-3d: High-fidelity 3d creation from a single image with diffusion prior,'' {\em arXiv preprint arXiv:2303.14184}   (2023).

\bibitem{goodfellow2020generative}
I.~Goodfellow, J.~Pouget-Abadie, M.~Mirza, {\em et~al.}, ``Generative adversarial networks,'' {\em Communications of the ACM} {\bf 63}(11), 139--144  (2020).

\bibitem{metz2016unrolled}
L.~Metz, B.~Poole, D.~Pfau, {\em et~al.}, ``Unrolled generative adversarial networks,'' {\em arXiv preprint arXiv:1611.02163}   (2016).

\bibitem{goodfellow2014generative}
I.~Goodfellow, J.~Pouget-Abadie, M.~Mirza, {\em et~al.}, ``Generative adversarial nets,'' {\em Advances in neural information processing systems} {\bf 27}  (2014).

\end{thebibliography}
\bibliographystyle{spiejour}   

\end{spacing}
\end{document}